\overfullrule=0pt
\input harvmac
\def\bar{\overline}
\def\a{{\alpha}}
\def\ad{{\dot a}}
\def\ah{{\widehat \a}}

\def\kb{{\bar \kappa}}

\def\l{{\lambda}}
\def\lb{{\bar\lambda}}

\def\lb{{\overline\lambda}}
\def\b{{\beta}}
\def\bh{{\widehat\beta}}
\def\dh{{\widehat\delta}}
\def\g{{\gamma}}
\def\gh{{\widehat\gamma}}

\def\d{{\delta}}
\def\e{{\epsilon}}
\def\s{{\sigma}}
\def\k{{\kappa}}
\def\kb{{\bar\kappa}}

\def\L{{\Lambda}}

\def\half{{1\over 2}}
\def\p{{\partial}}

\def\pb{{\overline\partial}}
\def\t{{\theta}}

\def\T{{\Theta}}
\def\S{{\Sigma}}

\def\Tb{{\bar\Theta}}
\def\tb{{\bar\theta}}

\Title{\vbox{\baselineskip12pt
\hbox{IFT-P.018/2007 }}}
{{\vbox{\centerline{ Towards a Worldsheet Derivation of the
 }
\smallskip
\centerline{Maldacena Conjecture}}} }
\bigskip\centerline{Nathan Berkovits\foot{e-mail: nberkovi@ift.unesp.br}}
\bigskip
\centerline{\it Instituto de F\'\i sica Te\'orica, State University of
S\~ao Paulo}
\centerline{\it Rua Pamplona 145, 01405-900, S\~ao Paulo, SP, Brasil}
\bigskip
\centerline{and}
\bigskip
\centerline{Cumrun Vafa\foot{e-mail: vafa@physics.harvard.edu}}
\bigskip
\centerline{\it Jefferson Laboratory, Harvard University}
\centerline{\it Cambridge, MA 02138, USA}

\vskip .3in

A $U(2,2|4)$-invariant A-model constructed from fermionic
superfields has recently been proposed as a sigma model
for the superstring on $AdS_5\times S^5$. After explaining the relation
of this A-model with the pure spinor formalism, the A-model action
is expressed as a gauged linear sigma model. In the zero radius limit,
the Coulomb branch of this sigma model is interpreted as D-brane holes
which are related to gauge-invariant
${\cal N}=4$ d=4 super-Yang-Mills operators. As in the
worldsheet derivation of open-closed duality for Chern-Simons theory, this
construction may lead to a worldsheet derivation of the Maldacena conjecture.
Intriguing connections to the twistorial formulation of ${\cal N}=4$ Yang-Mills
are also noted.

\vskip .3in

\Date {November 2007}

\newsec{Introduction}
Large $N$ dualities between gauge theories and gravity have been an important
development in our understanding of string theory.  In particular a large collection
of D-branes can be equivalently described by a dual purely gravitational system which
the D-branes generate.  A prominent example of this \ref\malda{J. Maldacena, {\it
The Large N Limit of Superconformal Field Theories and Supergravity}, Adv. Theor. Math. Phys.
2 (1998) 231, hep-th/9711200.}\
is the duality between the gauge system
living on $N$ D3 branes in the $\alpha'\rightarrow 0$ limit (i.e., ${\cal N}=4$ supersymmetric
$U(N)$ Yang-Mills in $d=4$) and the dual $AdS_5\times S^5$ where the D-branes have been
replaced with flux.

{}From the worldsheet perspective the duality can be interpreted as follows:  Let $\lambda$
denote the string coupling constant.  For each genus
$g$ in perturbation theory, on the D-brane side we have to insert an arbitrary number of holes
$h$ ending on the D-branes.  This gives rise to the factor $N^h$ for such amplitudes.
In addition this diagram is weighted with $\lambda^{2g-2+h}$.  Thus altogether we have a factor
$$F_{g,h} \lambda^{2g-2+h}N^h.$$
We consider first summing over 
the number of holes.  Replacing $N\lambda=T$, the 't Hooft parameter, we have
$$\lambda^{2g-2}\sum_h F_{g,h} T^h =\lambda^{2g-2} F_g$$
where
$$F_g(T)=\sum_h F_{g,h} T^h $$
is interpreted as the genus $g$ correction of a dual gravitational system where $T$ plays the
role of a modulus in the gravitational dual.  In other words the large $N$ duality is a statement
that can be seen order by order in closed string pertubation theory.  The subtlety is only
that the effective open string coupling $N\lambda=T$ can be large.  For large $T$ the gravitational
description is the better description and for small $T$ the gauge theory description, involving
D-branes.  

One idea for a 
perturbative proof of the
Maldacena conjecture would thus involve showing that if we start with the closed string description
of the system and take $T\rightarrow 0$ the worldsheet description will develop two phases (H,C), in one of
which (C) the degrees of freedom are frozen out.  Viewed from the perspective 
of the $H$ system we thus
have holes where the worldsheet is in the (C) phase.  One has to show that the amplitudes are non-vanishing
only if the (C) phase has the topology of a disc and that the path-integral
on each (C) region gives the correct factor of $N \lambda$.  This idea, which was suggested in \ref\gova{R. Gopakumar and 
C. Vafa,
{\it On the Gauge Theory/Geometry Correspondence}, 
Adv. Theor. Math. Phys. 3 (1999)
1415, hep-th/9811131.}\ 
in the context of large $N$ duality between $U(N)$ Chern-Simons theory on $S^3$
and resolved conifold geometry, was implemented in \ref\ova{H. Ooguri and C. Vafa,
{\it Worldsheet Derivation of a Large N Duality}, Nucl. Phys. B641 (2002) 3, hep-th/0205297.}\ and also
applied to derivations of duality for the F-terms in its superstring embedding \ref\bov{
N. Berkovits, H. Ooguri and C. Vafa, {\it On the Worldsheet Derivation of Large N Dualities for
the Superstring}, Comm. Math. Phys. 252 (2004) 259, hep-th/0310118.}.
Prominent in this derivation was the rewriting of a topological A-model in the form of a linear
sigma model and identifying the two phases as $H=Higgs$ and $C=Coulomb$ branches of the
sigma model as the modulus $T$ of the closed string approaches zero.
The aim of the present paper is to propose a similar scenario for the large $N$ duality
of ${\cal N}=4$ Yang-Mills and $AdS_5\times S^5$.

A basic step in this direction has already been taken \ref\limitone{N. Berkovits,
{\it New Limit of the $AdS_5\times S^5$ Sigma Model}, JHEP 0708 (2007) 011, hep-th/0703282.}.
In particular
it was shown that the gravity side, i.e. type IIB superstrings on $AdS_5\times S^5$ geometry,
can be viewed as an A-model topological string on the coset ${U(2,2|4)}\over{U(2,2)\times U(4)}$.
Here we make this map more precise and furthermore recast it as a gauged
linear sigma model.  In this
formulation, as the closed string modulus approaches zero, once again we obtain
two branches.   We will argue, just as in the Chern-Simons case, that the Coulomb branch
corresponds to holes in this formulation.  We thus end up with a worldsheet 
with an arbitrary
number of holes, which can then be interpreted as the `t Hooft diagrams of ${\cal N}=4$ supersymmetric
$U(N)$ Yang-Mills theory.  As evidence for this derivation we show how the half BPS sector of
the two sides map to one another in this setup.
In addition we find an intriguing connection with the twistorial formulation of ${\cal N}=4$
Yang-Mills:  a generic point on the Coulomb branch of the linear sigma model gives
four copies of ${\bf CP}^{3|4}$.  Even though we do not exploit this connection it is rather
suggestive.

The organization of this paper is as follows:  In section 2 we review the A-model formulation
of the $AdS_5\times S^5$.  In section 3, the relation between this A-model formulation
and the pure spinor fomulation of $AdS_5\times S^5$ is clarified.  In section 4 we review
the derivation
of the large $N$ duality between Chern-Simons and topological strings on the resolved
conifold.  In section 5 we construct the gauged linear sigma model
and propose a large $N$ derivation for our sigma model.  In section
6 we discuss our conclusions and open questions.

\newsec{Review of A-Model}

\subsec{Worldsheet variables}

In \limitone,
an N=2
worldsheet supersymmetric A-model was conjectured to describe the
superstring on $AdS_5\times S^5$.
Instead of being constructed using the ${PSU(2,2|4)}\over{SO(4,1)\times SO(5)}$
supercoset of Metsaev-Tseytlin, the variables in the A-model
are described by N=2 worldsheet superfields whose lowest components
take values in the supercoset ${U(2,2|4)}\over{U(2,2)\times U(4)}.$
(which can also be expressed as
${PU(2,2|4)}\over
{SU(2,2)\times U(4)}$ or 
${PSU(2,2|4)}\over
{SU(2,2)\times SU(4)}$). Since this supercoset only has fermionic elements,
the worldsheet superfields are all fermionic and will be called
$\T^A_J$ and $\Tb^J_A$ where $A=1$ to 4 and $J=1$ to 4 label
fundamental representations of $U(2,2)$ and $U(4)$ respectively. Furthermore,
$\T^A_J$ and $\Tb^J_A$ will be defined to be N=2 chiral and antichiral
superfields with the component expansions
\eqn\chir{\T^A_J(\k_+,\k_-) 
= \t^A_J + \k_+ Z^A_J + \k_- \bar Y^A_J + \k_+\k_- f^A_J,}
$$\Tb^J_A(\kb_+,\kb_-) 
= \tb^J_A + \kb_+ \bar Z^J_A + \kb_- Y^J_A + \kb_+\kb_- \bar f^J_A,$$
where $(\k_+,\kb_+)$ are left-moving and
$(\k_-,\kb_-)$ are right-moving Grassmann parameters, and $f^A_J$ and
$\bar f^J_A$ are auxiliary fields.  

As discussed in 
\limitone,
the 32 variables $\t^A_J$ and $\tb_A^J$ are related
to the usual 32 fermionic variables of $AdS_5\times S^5$ superspace,
whereas the 32 bosonic variables $Z^A_J$ and $\bar Z_A^J$ are
twistor-like combinations of the 10 spacetime variables $x^M$ and
the 22 pure spinor ghost variables $(\l^\a,\lb^\ah)$ of the pure
spinor formalism. Note that $d=10$ spacetime vectors will be denoted using
either the index $M=0$ to 9 or the $AdS_5\times S^5$ indices
$(m,\tilde m)=1$ to 5. And $d=10$ spacetime spinors will be denoted using
either the index $\a=1$ to 16 or $\ah=1$ to 16 depending if, in
a flat background, the spacetime
spinors are left or right-moving on the worldsheet.

To express $Z^A_J$ and $\bar Z_A^J$ in terms of $(x,\l,\lb)$,
first parameterize 
the $AdS_5$ variable
$x^m$ for $m=1$ to 5
as an ${SU(2,2)}\over {SO(4,1)}$ coset $H^A_{A'}(x)$
where $A'=1$ to 4 is an $SO(4,1)$ spinor index,
and parameterize 
the $S^5$ variable
$\tilde x^{\tilde m}$ for $\tilde m=1$ to 5 
as an ${SU(4)}\over {SO(5)}$ coset $\tilde H^J_{J'}(\tilde x)$
where $J'=1$ to 4
is an $SO(5)$ spinor index. Writing
the $SO(9,1)$ spinor index in terms of these $SO(4,1)\times SO(5)$
spinor indices, the left and right-moving pure spinor variables 
$\l^\a$ and $\lb^\ah$ satisfying $\l\g^M\l=0$ and $\lb\g^M\lb=0$ decompose as
$\l^{A'}_{ J'}$ and $\lb_{A'}^{ J'}$ which satisfy
\eqn\fivec{\l^{A'}_{ J'} \s^m_{A' B'}(\tilde\s^6)^{J'K'}\l^{B'}_{K'}=
\l^{A'}_{ J'}\s^6_{A'B'} (\tilde\s^{\tilde m})^{J' K'}\l^{B'}_{K'}=0,}
$$\lb_{A'}^{ J'} (\s^m)^{A' B'}\tilde\s^6_{J'K'}\lb_{B'}^{K'}=
\lb_{A'}^{ J'} (\s^6)^{A'B'}\tilde\s^{\tilde m}_{J' K'}\lb_{B'}^{K'}=0,$$
where $(\s^m_{A' B'}, \s^6_{A' B'})$ are the six Pauli matrices for 
$SO(4,2)=SU(2,2)$
and $(\tilde\s^{\tilde m}_{J' K'}, \tilde\s^6_{J' K'})$ are the six 
Pauli matrices for $SO(6) =SU(4)$. Note that $SO(4,1)\times SO(5)$
spinor indices can be raised and lowered using $\s^6_{A'B'}$ and
$\tilde\s^6_{J'K'}$, however, it will be convenient to always write
$\l^\a$ and $\lb^\ah$ as $\l^{A'}_{J'}$ and $\lb_{A'}^{J'}$.

The twistor-like combinations $Z^A_J$ and $\bar Z^J_A$ are constructed
from these $x$'s and $\l$'s as
\eqn\Zrel{Z^A_J = H^A_{A'}(x) (\tilde H^{-1}(\tilde x))_J^{J'} \l^{A'}_{J'},
\quad
\bar Z^J_A = (H^{-1}(x))^{A'}_A \tilde H^J_{J'}(\tilde x)\lb_{A'}^{J'}.}
Since $(x^m,\tilde x^{\tilde m})$ and $(\l^\a,\lb^\ah)$ contain
32 independent components and since $Z^A_J$ and $\bar Z^J_A$ are
unconstrained, the construction of \Zrel\ is invertible for
generic values of $(x,\tilde x)$ and $(\l,\lb)$. 
So for generic values of $(Z^A_J,\bar Z^J_A)$, the inverse map of \Zrel\
gives a point $(x^m,\tilde x^{\tilde m})$ on $AdS_5\times S^5$ together
with a pair of pure spinors $(\l^\a,\lb^\ah)$. In d=10 Euclidean space,
one can treat $\lb_\a\equiv (\g^{01234})_{\a\ah}\lb^\ah$ as the complex
conjugate of $\l^\a$, which implies that $\bar Z^J_A$ is the complex 
conjugate of $Z^A_J$.

\subsec{Worldsheet action}

As discussed in \limitone,
the $U(2,2|4)$-invariant action for the A-model can be written
in N=(2,2) superspace as
\eqn\sone{S = t\int d^2 z\int d^4\k ~Tr [\log (\d_K^J+\Tb^J_A \T^A_K)]}
where $t$ is a constant parameter and the notation $\log(M_K^J)$ denotes
the matrix $(\log M)_K^J$. The bosonic $U(2,2)\times U(4) $
isometries act in the obvious way as 
\eqn\transf{\d\T^A_J = i\L^A_B \T^B_J + i\tilde\L^K_J \T_K^A,\quad
\d\Tb^J_A = -i\L_A^B \Tb_B^J - i\tilde\L_K^J \Tb^K_A,}
and the 32 fermionic isometries act nonlinearly as
\eqn\transfsu{\d\T^A_J = \e^A_J + \T^A_K\bar\e_B^K \T^B_J,\quad
\d\Tb^J_A = \bar\e_A^J + \Tb_B^J\e_K^B \Tb_A^K.}
One can easily check that under the fermionic isometries,
$\d Tr[\log( \d_K^J+\Tb^J_A \T^A_K)] = \bar\e^J_A\T^A_J + \Tb_A^J\e^A_J$,
and since $\T^A_J$ and $\Tb_A^J$ are chiral and antichiral, the
action of \sone\ is invariant.

After integrating out the auxiliary fields $f^A_J$ and $\bar f_A^J$, the
action of \sone\ can be written in terms of the component fields of
\chir\ as 
\eqn\stwo{S= t\int d^2 z [ (G^{-1}\p G)^A_J (G^{-1}\pb G)_A^J}
$$
-Y^J_A (\bar\nabla  Z)^A_J + 
\bar Y_J^A (\nabla \bar Z)_A^J + (YZ)^J_K (\bar Z\bar Y)^K_J
- (ZY)^A_B (\bar Y\bar Z)_A^B] $$
where $G(\t,\tb)$ takes values in the fermionic coset ${U(2,2|4)}\over
{U(2,2)\times U(4)}$ which has 32 fermionic parameters, 
$(G^{-1}\p G)$ and $(G^{-1}\pb G)$ are the left-invariant
currents
taking values in the Lie algebra of $U(2,2|4)$,
$(YZ)^J_K = Y^J_A Z^A_K$, $(ZY)^A_B= Z^A_K Y^K_B$, and
\eqn\defnab{(\bar\nabla Z)^A_J = \pb Z^A_J + (G^{-1}\pb G)^A_B Z^B_J
- (G^{-1}\pb G)^K_J Z^A_K,}
$$
(\nabla \bar Z)^J_A = \p \bar Z^J_A - (G^{-1}\p G)^B_A \bar Z_B^J
+ (G^{-1}\pb G)_K^J \bar Z_A^K.$$

Note that N=(2,2) worldsheet supersymmetry is manifest using the superspace
form of the action of \sone,
whereas $U(2,2|4)$ symmetry is manifest using the
component form of the action of \stwo. 
As will be shown in section 5, both these
symmetries can be made manifest by writing the action as a gauged linear
sigma model. Furthermore, it was shown in \limitone\ that this A-model
action has no conformal anomaly.

\subsec{Open string sector}

As discussed in \limitone, a natural open string boundary condition for
the A-model is
\eqn\osb{\Tb_A^J = \d^{JK}\e_{AB}\T^B_K}
where $\e_{AB}$ is an antisymmetric tensor which breaks $SU(2,2)$ to
$SO(3,2)$.
The boundary condition of \osb\ is similar to the open
string boundary condition for the Chern-Simons topological string which is
$\bar X_I = \d_{IJ} X^J$ for $I,J=1$ to 3. Note that the open string
boundary for the A-model is defined by
$z=\bar z$, $\k_+ = \bar \k_-$, and $\kb_+ = \k_-$,
so \osb\ implies that
\eqn\osbtwo{\tb^J_A = \d^{JK}\e_{AB}\t^B_K, \quad
\bar Z^J_A = \d^{JK}\e_{AB}Z^B_K,\quad
Y^J_A = \d^{JK}\e_{AB}\bar Y^B_K.}
The boundary condition of \osb\ breaks half of the fermionic isometries
and reduces the $U(2,2|4)$ supergroup of isometries
to the supergroup $OSp(4|4)$. This supergroup contains
$SO(3,2)\times SO(4)$ bosonic isometries and 16 fermionic isometries,
and is the ${\cal N}=4$ supersymmetry algebra on $AdS_4$.

In \limitone, it was conjectured that the open string sector of the A-model
might describe ${\cal N}=4$ d=4 super-Yang-Mills in the same manner that
the open sector of Witten's topological A-model describes $d=3$
Chern-Simons. Evidence for this conjecture came from the fact that the
$\a'\to 0$ limit of this open string sector is described by the
pure spinor superparticle whose spectrum is ${\cal N}=4$ d=4 super-Yang-Mills.
However, it was not proven that there are no massive states in the
open string sector coming from the worldsheet nonzero modes.

In this paper, the conjecture that the open string sector of the A-model
contains only 
${\cal N}=4$ d=4 super-Yang-Mills states will be withdrawn, and it will instead
be argued
that the open string boundary conditions of \osb\ describe an $AdS_4$
D-brane probe embedded in $AdS_5\times S^5$. Although the low-energy
sector of this D-brane probe contains ${\cal N}=4$ d=4 super-Yang-Mills states,
one also expects to have massive states in the spectrum. Note that
the position of this $AdS_4$ D-brane probe in $AdS_5$ is determined
by the choice of the antisymmetric tensor $\e_{AB}$ in \osb. 
There are ${SO(4,2)}\over{SO(3,2})$ different ways to embed $AdS_4$
in $AdS_5$, and the choice of $\e_{AB}$ determines this embedding.

\newsec{Relation of A-model with Pure Spinor Formalism}

In this section, the relation between the A-model action of \sone\
and the pure spinor $AdS_5\times S^5$ sigma model will be clarified.
(In \limitone, the relation between these actions was understood
only in a certain singular limit of the superspace torsion.)
Using the field redefinition of \Zrel, it will be shown that the A-model
maps into the pure spinor sigma model where the parameter
$t$ in \sone\ is related to the
$AdS_5$ radius $R$ as $t=\half R^2$.
When $t\to \infty$, the A-model becomes weakly coupled and describes the
flat-space limit of the $AdS_5\times S^5$ sigma model. And when $t\to 0$,
the A-model becomes strongly coupled and describes the highly curved limit
of the $AdS_5\times S^5$ sigma model. As will be discussed
in section 5, much can be learned about the $t\to 0$ limit by writing
the A-model as a gauged linear sigma model.

Although the A-model of \sone\ is invariant under $U(2,2|4)$ global
isometry, the pure spinor $AdS_5\times S^5$ sigma model (like the
Green-Schwarz $AdS_5\times S^5$ sigma model) is only invariant under
$PSU(2,2|4)$ isometry. Nevertheless, it will be shown in subsection (3.1)
that after adding a BRST-trivial term, the pure spinor sigma model can
be expressed as a $U(2,2|4)$-invariant action. 
The field redefinition of \Zrel\ will then be used in subsection (3.2) to map
this $U(2,2|4)$ invariant form of the pure spinor sigma model 
into the A-model action of \sone.

Since the physical theory described by the sigma model is invariant
under only $PSU(2,2|4)$ isometry, a natural question is how the bonus
U(1) symmetry is broken. (Note that one of the $U(1)$'s in $U(2,2|4)$
acts trivially on all fields. The ``bonus'' $U(1)$ is the
symmetry in $PU(2,2|4)$ which is not in $PSU(2,2|4)$.) 
As will be discussed in subsection
(3.3), the bonus $U(1)$ symmetry is preserved by the worldsheet action but
will be broken by the BRST operator which determines the physical state
conditions.

\subsec{$U(2,2|4)$-invariant pure spinor sigma model}

Using the conventions of \ref\adsq{N. Berkovits,
{\it Quantum Consistency of the Superstring in $AdS_5\times S^5$ Background}, 
JHEP 0503 (2005) 041,
hep-th/0411170.}, the pure spinor sigma model action is
\eqn\sigmapure{S = R^2 \int d^2 z [\half \eta_{MN} J^M \bar J^N
- \eta_{\a\bh}({3\over 4} J^\bh \bar J^\a + {1\over 4} \bar J^\bh J^\a)}
$$- w_\a \bar\nabla\l^\a  
+ \bar w_\ah \nabla\lb^\ah 
-{1\over 4} \eta_{[MN][PQ]} (w\g^{[MN]}\l)(\bar w \g^{[PQ]}\lb)]$$
where
$$\bar \nabla\l^\a = (\pb\l + \half J^{[MN]}\g_{MN}\l)^\a, \quad
\nabla\lb^\ah = 
(\p\lb + \bar J^{[MN]}\g_{[MN]}\lb)^\ah,$$ 
and $J=( g^{-1}\p g)$ and
$\bar J= (g^{-1}\pb g)$ are the Metsaev-Tseytlin left-invariant 
currents constructed from a matrix $g(x,\t,\tb)$ taking values in
the supercoset ${PSU(2,2|4)}\over{SO(4,1)\times SO(5)}$. These currents
$J$ take values in the 
$PSU(2,2|4)$ Lie algebra
where $J^M = (J^m,J^{\tilde m})$ are the 10 translation
currents, $J^\a$ and $J^\ah$ are the 32 supersymmetry currents,
and $J^{[MN]} = (J^{[mn]},J^{[\tilde m\tilde n]})$ are
the 20
$SO(4,1)\times SO(5)$ Lorentz currents. 
Furthermore, $\eta_{\a\bh} = (\g^{01234})_{\a\bh}$, 
$\eta_{[mn][pq]} = \eta_{m[p}\eta_{q]n}$ and
$\eta_{[\tilde m\tilde n][\tilde p\tilde q]} = 
-\eta_{\tilde m[\tilde p}\eta_{\tilde q]\tilde n}$.

Under the ``bonus'' $U(1)$ symmetry of $PU(2,2|4)$, $J^\a$ and $J^\ah$
rotate into each other as
\eqn\rotatej{\d J^\a = i J^\ah,\quad \d J^\ah = -i J^\a.}
In other words, $(J^\a \pm i J^\ah)$ carries $\pm$ $U(1)$ charge under
this symmetry. Since $J^M$ and $J^{[MN]}$ are $U(1)$ invariant, the
action of \sigmapure\ transforms under the bonus $U(1)$ as
\eqn\transu{\d S = R^2 \int d^2 z [ -i\eta_{\a\b} J^\b \bar J^\a + i
\eta_{\ah\bh} J^\bh \bar J^\ah ]} 
where $\eta_{\a\b} = (\g^{01234})_{\a\b}$ and 
$\eta_{\ah\bh} = (\g^{01234})_{\ah\bh}$.
Nevertheless, by adding a BRST-trivial term to the action, 
this $U(1)$ transformation
can be cancelled. The resulting $U(2,2|4)$-invariant action can then
be mapped into the A-model action of \sone.

The BRST-trivial term is given by
\eqn\sbt{S_{trivial} = -\half R^2 \int d^2 z [\half {
{\eta^{\a\bh} (\g^M\l)_\a (\g^N \lb)_\bh}
\over{(\eta\l\lb)}}
J^M \bar J^N -\eta_{\a\bh} J^\bh \bar J^\a }
$$- w_\a \bar\nabla\l^\a  
+ \bar w_\ah \nabla\lb^\ah 
-{1\over 4} \eta_{[MN][PQ]} (w\g^{[MN]}\l)(\bar w \g^{[PQ]}\lb)].$$
Note that the second line of \sbt\ is identical to the second line of
\sigmapure\ whose BRST transformation under 
\eqn\brstold{Q+\bar Q = \int dz \eta_{\a\bh} \l^\a J^\bh + \int d\bar z
\eta_{\a\bh} \lb^\bh \bar J^\a}
is 
\eqn\brsttra{\eta_{\a\bh} (- J^\bh \bar\nabla \l^\a + 
\bar J^\a \nabla\lb^\bh).}
Using the transformations
\eqn\brsttww{(Q+\bar Q) J^\a = \nabla \l^\a - 
\eta^{\a\bh} (\g_M\lb)_\bh J^M,}
$$(Q+\bar Q) J^M = J^\a (\g^M \l)_\a + J^\ah (\g^M \lb)_\ah,$$
$$
(Q+\bar Q) J^\bh = \nabla \lb^\bh + \eta^{\a\bh} (\g_M\l)_\a J^M,$$ 
and the identity
$\g^M_{\ah\bh} = \eta_{\a\ah} \eta_{\b\bh} (\g^M)^{\a\b}$,
it is easy to verify that the BRST transformation of the first line of 
\sbt\ cancels \brsttra, so that $S_{trivial}$ is BRST-closed. 
Furthermore, the coefficient $-\half R^2$
multiplying $S_{trivial}$ has been chosen so that the bonus $U(1)$
transformation of $S_{trivial}$ cancels \transu.

Finally, one can show that $S_{trivial}$ is BRST-trivial by writing it
as $S_{trivial} = Q \bar Q \Omega$ where
\eqn\sbtw{\Omega = -\half R^2 \int d^2 z {1\over{(\eta\l\lb)}}
[{1\over 4}(w\l)(\bar w\lb) -{1\over 8} (w\g^{MN}\l)(\bar w\g_{MN}\lb) 
+{1\over 4} 
{{\eta^{\a\bh} (\g^M\l)_\a (\g^N \lb)_\bh}
\over{(\eta\l\lb)}} J^M \bar J^N].}
To show that $S_{trivial} = Q \bar Q \Omega$, one uses the identity
\eqn\iden{\d_\b^\g \d_\a^\d = \half \g^M_{\a\b} \g_M^{\g\d}
-{1\over 8} (\g^{MN})_\a^\g (\g_{MN})_\b^\d -{1\over 4} \d_\a^\g \d_\b^\d,}
together with the BRST transformations of \brsttww\ and
\eqn\brstthr{Q w_\a = \eta_{\a\ah} J^\ah, \quad \bar Q w_\a = w^*_\a,}
$$Q \bar w_\ah = \bar w^*_\ah,\quad
\bar Q \bar w_\ah = \eta_{\a\ah} J^\a, $$
$$Q w_\a^* = \eta_{\a\ah} (\nabla \lb^\ah -{1\over 4} \eta_{[MN][PQ]}
(w\g^{[MN]}\l)(\g^{[PQ]}\lb)^\ah), \quad \bar Q w_\a^*=0,$$
$$Q\bar w_\ah^* =0,\quad
\bar Q \bar w_\ah^* = \eta_{\a\ah} (\bar\nabla \lb^\a +{1\over 4} 
\eta_{[MN][PQ]} (\g^{[MN]}\l)^\a(\bar w\g^{[PQ]}\lb)).$$

In reference \adsq, the auxiliary variables $w_\a^*$ and $\bar w_\ah^*$ 
were not included, and the BRST transformations were nilpotent only up
to the equations of motion 
$$\nabla \lb^\ah -{1\over 4} \eta_{[MN][PQ]}
(w\g^{[MN]}\l)(\g^{[PQ]}\lb)^\ah =0, \quad
\bar\nabla \lb^\a +{1\over 4} 
\eta_{[MN][PQ]} (\g^{[MN]}\l)^\a(\bar w\g^{[PQ]}\lb)=0.$$
Note that $Q w_\a^*$ and $\bar Q \bar w_\ah^*$ are proportional to
these equations of motion which come from varying $w_\a$ and $\bar w_\ah$.
So after adding the term 
\eqn\adda{R^2\int d^2 z \eta^{\a\bh} \bar w_\bh^* w_\a^*}
to the pure spinor sigma model action of \sigmapure, the action will
be invariant with respect to the BRST transformations of \brstthr.
The auxiliary variables $w_\a^*$ and $\bar w_\ah^*$
can be naturally
interpreted as antifields which allow the BRST transformation generated
by $(Q+\bar Q)$
to be nilpotent off-shell.\foot{The structure of the antifields $w_\a^*$
and $\bar w_\ah^*$ in the pure spinor $AdS_5\times S^5$ sigma model was
also discussed in independent work by Guillaume Boussard \ref\bous{
G. Boussard, private communication.}.}

In this construction of a $U(2,2|4)$-invariant pure spinor sigma
model, 
the only subtlety is the presence of inverse powers
of $(\eta_{\a\bh}\l^\a\lb^\bh)$ in $S_{trivial}$ and $\Omega$. If one
Wick-rotates both the $d=2$ and $d=10$ metric to Euclidean space, 
it is natural to define $\lb_\a \equiv\eta_{\a\bh}\lb^\bh$ to be
the complex conjugate of $\l^\a$.
Using this definition of complex
conjugation, $\l^\a \lb_\a$ is only zero if each component
of $\l^\a$ is zero. Therefore, $S_{trivial}$ and $\Omega$ are well-defined
except where $\l^\a = \lb_\a=0$. As in \ref\nekr{N. Nekrasov, {\it Lectures on Curved Beta-Gamma
Systems}, hep-th/0511008.} , we shall assume that we can
remove the singular point $\l^\a = 0$ from the pure spinor space so
that $S_{trivial}$ and $\Omega$ are well-defined.

One possible problem with removing the point $\l^\a=0$ is that, 
in a flat background, allowing operators such as 
\eqn\opeq{\xi = {{\t^\a \lb_\a}\over{(\eta\l\lb)}} }
in the Hilbert space implies that the BRST cohomology is trivial.
Since $Q\xi = 1$, any operator $V$ satisfying $QV=0$ can be written
as $V = Q(\xi V)$. However, $\xi$ of \opeq\
is not spacetime supersymmetric, and it was conjectured in \ref\relat
{N. Berkovits, {\it Relating the RNS and Pure Spinor Formalisms
for the Superstring}, JHEP 0108 (2001) 026, hep-th/0104247.} that
if one restricts operators with poles in $\l^\a$ to spacetime
supersymmetric operators (such as the composite $b$ ghost), these operators
do not trivialize the Hilbert space.

In the case of an $AdS_5\times S^5$ background, one can make a similar
conjecture with spacetime supersymmetric operators being replaced by
$PSU(2,2|4)$-invariant operators. Since $S_{trivial}$ and $\Omega$
are $PSU(2,2|4)$-invariant, the conjecture would imply these operators
do not cause problems.
Nevertheless, this subtlety certainly deserves further
investigation. 

\subsec{Mapping to the A-model}

After adding $S_{trivial}$ to the pure spinor sigma model action of
\sigmapure, one obtains the $U(2,2|4)$-invariant action
\eqn\sigmainv{S= \half R^2 \int d^2 z [\half 
{ {\eta^{\a\bh} (\g^M\l)_\a (\g^N \lb)_\bh}
\over{(\eta\l\lb)}}
\bar J^M J^N -\half\eta_{\a\bh} (J^\bh \bar J^\a +\bar J^\bh J^\a)}
$$- w_\a \bar\nabla\l^\a +
 \bar w_\a \nabla\lb^\ah 
-{1\over 4}\eta_{[MN][PQ]} (w\g^{[MN]}\l)(\bar w \g^{[PQ]}\lb)].$$
It will now be shown that this action is equivalent to the 
A-model action of \stwo\ where $t=\half R^2$.

The first step in relating the actions of \sigmainv\
and \stwo\ is to express the supercoset
$g\in {{PSU(2,2|4)}\over{SO(4,1)\times SO(5)}}$ in terms of the 
fermionic coset
$G\in 
{{U(2,2|4)}\over{U(2,2)\times U(4)}}$ and the bosonic variables
$Z^A_J$ and $\bar Z^J_A$. Using the definitions of \Zrel\ that
\eqn\Zrelt{Z^A_J = H^A_{A'}(x) (\tilde H^{-1}(\tilde x))_J^{J'} \l^{A'}_{J'},
\bar Z^J_A = (H^{-1}(x))^{A'}_A \tilde H^J_{J'}(\tilde x)\lb_{A'}^{J'}}
where $H^A_{A'}(x)\in {{SU(2,2)}\over {SO(4,1)}}$ and
$H^J_{J'}(\tilde x)\in {{SU(4)}\over {SO(5)}}$, it is natural
to parameterize $g$ as
\eqn\mapg{g(x,\tilde x, \t, \tb) = e^{\t^\a Q_\a + \tb^\a \bar Q_\a}
e^{x^m P_m} e^{\tilde x^{\tilde m }P_{\tilde m}} = 
G(\t,\tb) H(x)\tilde H(\tilde x)}
where $G(\t,\tb) = e^{\t^\a Q_\a + \tb^\a \bar Q_\a}$, $H(x)=
e^{x^m P_m}$, $\tilde H(\tilde x) =  e^{\tilde x^{\tilde m }P_{\tilde m}}$,
and $(P_m, P_{\tilde m}, Q_\a, \bar Q_\a)$ are the 10 translation and
32 supersymmetry generators on $AdS_5\times S^5$.

The map of \mapg\ implies that the left-invariant currents $J= g^{-1}\p g$
which appear in the pure spinor sigma model action are related to $G$ and $H$
as
\eqn\curre{J^{A'}_{B'} = (H^{-1} \p H)^{A'}_{B'} + (H^{-1})^{A'}_A
(G^{-1} \p G)^A_B H^B_{B'},}
$$J^{J'}_{K'} = (\tilde H^{-1} \p \tilde H)^{J'}_{K'} + (\tilde H^{-1})^{J'}_J
(G^{-1} \p G)^J_K H^K_{K'},$$
$$J^{A'}_{J'} = (H^{-1})^{A'}_A
(G^{-1} \p G)^A_J \tilde H^J_{J'},$$
$$J^{J'}_{A'} = (\tilde H^{-1})^{J'}_J
(G^{-1} \p G)^J_A \tilde H^A_{A'}.$$
In terms of the left-invariant currents of \sigmapure, 
\eqn\onetof{J^{A'}_{B'} = \half J^m (\s_{m}\s_6)^{A'}_{B'} + \half
J^{[mn]} (\s_{m}\s_{n})^{A'}_{B'},}
$$J^{J'}_{K'} = \half J^{\tilde m} 
(\tilde\s_{\tilde m}\tilde\s_6)^{J'}_{K'} + \half
J^{[\tilde m\tilde n]} 
(\tilde \s_{\tilde m}\tilde \s_{\tilde n})^{J'}_{K'},$$
$$J^{A'}_{J'} = {1\over{\sqrt 2}}[J^\a (f_\a)^{A'}_{J'} + i
J^\ah (f_\ah)_{B'}^{K'} (\s^6)^{A'B'} \tilde\s^6_{J'K'}],$$
$$J^{J'}_{A'} = {1\over{\sqrt 2}}[J^\ah (f_\ah)^{J'}_{A'} + i
J^\a (f_\a)^{B'}_{K'} \s^6_{A'B'} (\tilde\s^6)^{J'K'}],$$
where $(f_\a)^{A'}_{ J'}$ and
$(f_\ah)^{J'}_{ A'}$ are 
Clebsch-Gordon coefficients for decomposing
an $SO(9,1)$ spinor into an $SO(4,1)\times SO(5)$ spinor. 
Note that $J^{A'}_{J'}$ has bonus $U(1)$ charge $+1$ and
$J^{J'}_{A'}$ has bonus $U(1)$ charge $-1$, which explains the relative
coefficients in \onetof.

The next step in relating the two actions is to use the definitions of
$Z^A_J$ and $\bar Z_A^J$ in \Zrelt, together with the definitions
\eqn\Zrelt{
Y^J_A = (H^{-1}(x))^{A'}_A \tilde H^J_{J'}(\tilde x) w_{A'}^{J'},\quad
\bar Y^A_J = H^A_{A'}(x) (\tilde H^{-1}(\tilde x))_J^{J'} \bar w^{A'}_{J'},}
to relate the second lines of \sigmainv\ and \stwo. 
Since 
\eqn\firsteq{Y^J_A \pb Z^A_J = w_{A'}^{ J'}\pb \l^{A'}_{ J'} + 
(H^{-1} \bar\p H)^{A'}_{B'} w_{A'}^{ J'}\l^{B'}_{ J'} -
(\tilde H^{-1}\bar\p\tilde H)^{K'}_{J'} w_{A'}^{ J'}\l^{A'}_{ K'}}
$$=w_{A'}^{ J'}\pb \l^{A'}_{ J'} + \half (\bar J^m (\s_m\s_6)^{A'}_{B'}
+ \bar J^{[mn]}
(\s_m\s_n)^{A'}_{B'})(\l w)^{B'}_{A'} $$
$$- \half
(\bar J^{\tilde m}(\tilde\s_{\tilde m}\tilde\s_6)^{J'}_{K'} +
\bar J^{[\tilde m\tilde n]}
(\tilde\s_{\tilde m}\tilde\s_{\tilde n})^{J'}_{K'} )(w\l)^{K'}_{J'}
- (G^{-1}\pb G)^A_B (YZ)^B_A + (G^{-1}\pb G)^J_K (Y Z)^K_J,$$
one finds that 
\eqn\segeq{Y^J_A (\bar\nabla Z)^A_J =
w_\a (\pb\l + \half\bar J^{[MN]}\g_{[MN]}\l)^\a  +\half
\bar J^m (\s_m\s_6)^{A'}_{B'}
(\l w)_{A'}^{B'} 
-\half \bar J^{\tilde m} (\tilde\s_{\tilde m}\tilde \s_6)^{J'}_{K'} 
(w\l)^{K'}_{J'}.}
Similarly, one finds that 
\eqn\threq{\bar Y^A_J (\nabla \bar Z)^J_A =
\bar w_\a (\p\lb + \half J^{[MN]}\g_{[MN]}\lb)^\a 
-\half J^m (\s_m\s_6)^{A'}_{B'}(\bar w\lb)_{A'}^{B'}
+\half J^{\tilde m} (\tilde\s_{\tilde m}\tilde \s_6)^{J'}_{K'} 
(\lb \bar w)^{K'}_{J'},}
and that
\eqn\findot{(YZ)_K^J (\bar Z\bar Y)^K_J  -
(ZY)_B^A (\bar Y\bar Z)^B_A  = (w\l)^{J'}_{K'} (\lb \bar w)_{J'}^{K'} -
 (\l w)^{A'}_{B'} (\bar w\lb )_{A'}^{B'}.}

Putting \segeq - \findot\ together, one finds that the
A-model action of \stwo\ is equal to 
\eqn\nexts{S = t \int d^2 z [- {1\over 2}\eta_{\a\bh} (J^\bh \bar J^\a +
\bar J^\bh J^\a)}
$$- w_\a (\pb\l + \half J^{[MN]}\g_{[MN]}\l)^\a 
+ \bar w_\a (\p\lb + \half\bar J^{[MN]}\g_{[MN]}\lb)^\a 
-{1\over 4} \eta_{[MN][PQ]} (w\g^{[MN]}\l)(\bar w \g^{[PQ]}\lb)$$
$$+ \half\eta^{\a\b} w_\a (\g_M\l)_\b \bar J^M
+ \half\eta^{\ah\bh} \bar w_\ah (\g_M\lb)_\bh  J^M
- {1\over 4} (\eta^{\ah\bh} \bar w_\ah (\g_M\lb)_\bh)
(\eta^{\g\d} w_\g (\g^M\l)_\d)].$$

Under the transformation $\d w_\a = (\l\g^M)_\a \L_M$, the first two
lines of \nexts\ are invariant, but the last line transforms as
\eqn\deltr{\d S = t\int d^2 z ~\L^N \eta^{\a\b} (\g_N\l)_\a (\g_M\l)_\b
(\half\bar J^M - {1\over 4}\eta^{\gh\dh}\bar w_\gh (\g^M\lb)_\dh).}
Since $\d S=0$ onshell, one learns that the equations of motion for
$w_\a$ imply that
\deltr\ vanishes for any $\L^N$, which implies that 
\eqn\impleom{(\g_M\l)_\b (\bar J^M - \half
\eta^{\gh\dh}\bar w_\gh (\g^M\lb)_\dh)=0.}
Similarly, the equations of motion for $\bar w_\a$
imply that
\eqn\impleomb{(\g_M\lb)_\b (J^M - \half\eta^{\g\d}w_\g (\g^M\l)_\d)=0.}

Since the equations of motion of \impleom\ and \impleomb\ are auxiliary,
they can be plugged back into the action of \nexts. One finds that
all three terms in the last line of \nexts\ are proportional to each other,
and their sum is equal to 
\eqn\sumterms{t \int d^2 z \half
{{\eta^{\a\bh}(\g_M\l)_\a (\g_N\lb)_\bh}\over{(\eta\l\lb)}}
\bar J^M  J^N ,}
which coincides with the first term in \sigmainv.
So the A-model action of \stwo\ is equal to 
\eqn\finalact{
S= t \int d^2 z [\half 
{ {\eta^{\a\bh} (\g^M\l)_\a (\g^N \lb)_\bh}
\over{(\eta\l\lb)}}
\bar J^M J^N -\half\eta_{\a\bh} (J^\bh \bar J^\a +\bar J^\bh J^\a)}
$$- w_\a \bar\nabla\l^\a +
 \bar w_\a \nabla\lb^\ah 
-{1\over 4}\eta_{[MN][PQ]} (w\g^{[MN]}\l)(\bar w \g^{[PQ]}\lb)],$$
which coincides with the $U(2,2|4)$-invariant pure spinor sigma model
of \sigmainv\ when $t=\half R^2$.

\subsec{BRST operator}

For the A-model action of \sone\ and \stwo, the obvious guesses for 
left and right-moving BRST operators are the scalar generators
of N=2 worldsheet supersymmetry, 
\eqn\genetwo{\int dz Z^A_J (G^{-1}\p G)_A^J,\quad
\int d\bar z \bar Z_A^J (G^{-1}\pb G)_J^A.}
Surprisingly, these
do not match the left and right-moving
BRST operators in the pure spinor sigma model and would
therefore give the incorrect cohomology. 

Under the map of \mapg, it is easy to check that
\genetwo\ map into the operators
\eqn\mapwr{\int dz \l^\a (\eta_{\a\bh} J^\bh + i \eta_{\a\b} J^\b),\quad
\int d\bar z \lb^\bh (\eta_{\ah\bh}\bar J^\ah - i \eta_{\a\bh} J^\a).}
However, the pure spinor left and right-moving BRST operators are
\eqn\pureb{Q=\int dz \eta_{\a\bh} \l^\a J^\bh,\quad \bar Q=
\int d\bar z \eta_{\a\bh} \lb^\bh J^\a.}
So to reproduce the correct cohomology, one must 
map the left and right-moving
BRST operators of \pureb\ into the A-model variables, which implies 
\eqn\abrst{Q= \int dz Z^A_J [(G^{-1}\p G)^J_A + i
(H^{-1})^{A'}_A \tilde H^J_{J'} \s^6_{A' B'} (\tilde\s^6)^{J' K'}
(H^{-1})^{B'}_B \tilde H^K_{K'} (G^{-1}\p G)^B_K],}
$$\bar Q= \int d\bar z \bar Z^J_A [(G^{-1}\p G)^A_J + i
H_{A'}^A (\tilde H^{-1})_J^{J'} (\s^6)^{A' B'} \tilde\s^6_{J' K'}
H_{B'}^B (\tilde H^{-1})_K^{K'} (G^{-1}\p G)^K_B],$$
where $H$ and $\tilde H$ are defined in terms of 
$Z$ and $\bar Z$ by the inverse map
of \Zrel.

Note that after adding the BRST-trivial term of \sbt\ to the pure spinor
action, both 
\eqn\holo{\eta_{\a\bh} \l^\a J^\bh \quad{\rm and}
\quad \eta_{\a\b}\l^\a J^\b}
are holomorphic currents. This is easy to see since the action of
\sigmainv\ is $U(1)$ invariant, and $J^\ah$ and $J^\a$ transform
into each other under this $U(1)$. Furthermore, one can check that
the currents in \holo\ are nilpotent and satisfy the OPE
\eqn\opej{
(\eta_{\a\bh} \l^\a J^\bh(y)) ~~ (\eta_{\g\d}\l^\g J^\d(z)) \to (y-z)^{-2}
\eta_{\a\b} \l^\a(y) \l^\b (z).}
So the operators of
\holo\ satisfy the OPE's of the two spin-one fermionic generators,
$G^+$ and $\tilde G^+$, of a twisted ``small'' N=4 superconformal
algebra whose generators are 
$$[T, G^+,\tilde G^+, G^-, \tilde G^-,
J^{++}, J, J^{--}].$$ 

It is easy to explicitly construct the generators
\eqn\gplus{G^+ = \eta_{\a\bh} \l^\a J^\bh, \quad\tilde G^+=
\eta_{\a\b}\l^\a J^\b, \quad
J^{++} = \eta_{\a\b}\l^\a\l^\b, }
$$J = \l^\a w_\a,\quad
T = \half\eta_{MN} J^M J^N + \eta_{\a\bh} J^\bh J^\a - 
w_\a (\p\l +\half J^{[MN]}\g_{MN}\l)^\a,$$
however, the remaining N=4 generators do not appear to be easy to
construct. For example, the obvious guess for $J^{--}$ 
is $J^{--} = \eta^{\a\b}w_\a w_\b$, but this is not holomorphic.
Also, to construct $G^-$ and $\tilde G^-$, one
would need the analog of the composite $b$ ghost in the pure spinor
formalism which is not easy to construct even in a flat background.

Nevertheless, the existence of an ``almost'' N=4 superconformal
algebra for the A-model with the generators of \gplus\ 
allows the construction of the nilpotent left and right-moving
BRST operators of \abrst\ which differ from the naive guess of \genetwo. 
In other words, the existence of an ``almost'' N=4 algebra allows the
choice of 
\eqn\choiceb{Q = \int dz (A G^+ + B \tilde G^+)}
where $A$ and $B$ are arbitrary constants. The naive guess of 
\genetwo\ corresponds to $B=iA$, whereas the map of the pure spinor
BRST operators of \abrst\ corresponds to $B=0$. 

Since the bonus $U(1)$ symmetry is preserved only if $A^2+B^2=0$,
it seems reasonable to conjecture that any choice of the constants
$A$ and $B$ is allowed as long as $A^2+B^2$ is nonzero. If this
conjecture is correct, the cohomology of \choiceb\ should be 
independent of $A$ and $B$, except for the singular
choice where $A^2 + B^2 =0$.

\newsec{Worldsheet derivation of Chern-Simons/topological gravity duality}

In this section we will review the worldsheet derivation of the duality between the A-model
topological string on the resolved conifold and Chern-Simons $U(N)$ gauge theory on $S^3$ \ova,
along the lines proposed in \gova .

The basic idea is to start from the closed string side, i.e. the
topological A-model on the resolved
conifold, with $t$ being the modulus of ${\bf P}^1$.  One then considers expanding the A-model
sigma model near the $t\rightarrow 0$ limit, as a perturbation in $t$.  However, the non-linear
sigma model is singular in this limit. Instead one considers the gauged linear sigma model formulation
of the conifold where the Higgs branch of the $U(1)$ gauge system flows in the IR to the
non-linear sigma model of the resolved conifold. The gauged linear sigma model
is {\it not} singular in this regime, and the fact 
that the geometry is singular translates in this formulation to the opening up of a new branch for the
%NB please check the above rephrasing is correct
gauge theory:  the Coulomb branch \ref\silwi{E. Silverstein and E. Witten,
{\it Criteria for Conformal Invariance of (0,2) Models}, Nucl. Phys. B444 (1995) 161,
hep-th/9503212.}.  In other words the operation of going to the IR and going to the
non-linear sigma model formulation do not commute in this limit and we end up with a completion of
the worldsheet theory using the gauged linear sigma model.
 
In the limit as $t\rightarrow 0$ the  path-integral of the worldsheet theory will have two regions.
One region is in the Higgs branch and the other in the Coulomb branch.  Moreover
the degrees of freedom which are light in the Higgs branch are massive in the Coulomb branch.  In the UV the separation between
these regions is not sharp.  But as we take the IR limit the separation becomes sharper.  Indeed
we have to take the IR limit as the CFT description of the worldsheet is what arises in string
perturbation theory.
 
We thus end up with a worldsheet marked by $H$ and $C$ regions.  However, it turns out that
if the topology of the $C$ region is anything but a disc, then the amplitude vanishes \ova .  This
is because the contribution of each $C$ region to the path integral ends up being a total derivative
in the moduli.  In particular if $\theta$ denotes the angular part of $t$, the contribution of
each $C$ region is given as
$$\oint d\theta {\partial F\over \partial \theta}$$
This is manifestly zero except when $F$ is not a single valued function of $\theta$. This is only
the case for the disc topology where the partition function of a topological string is not well defined
(due to the $SL(2,R)$ symmetry)\foot{There are potentially a few other possibilities, but these
can
be ruled out \ova .}.  We thus end up with the Higgs branch and a number of Coulomb branches, all of
which are in disc topology.  Moreover, since the fields in the Higgs branch are massive in the Coulomb
branch, in the IR they vanish as they approach the Coulomb branch. In other words the discs play
the role of holes with D-brane boundary condition for the Higgs branch.  Moreover one can show that
up to a BRST trivial deformation (deforming the $t=0$ conifold to the deformed conifold) these
can be viewed as Lagrangian D-branes living on $S^3$.  Moreover doing the path-integral
on each Coulomb branch yields a factor of $t$ which is identified with $t=N\lambda$.  In particular
this is the right structure for it to correspond to $N$ Lagrangian D-brane insertions on $S^3$.
This makes contact with Witten's formulation of Chern-Simons theory as 
an A-model with $N$ D-branes
wrapping Lagrangian submanifolds \ref\wittcs{E. Witten, {\it Chern-Simons Gauge Theory
as a String Theory}, Prog. Math. 133 (1995) 637, hep-th/9207094.}.  
We thus end up with a description
of the theory in terms of a $U(N)$ Chern-Simons gauge theory.

\subsec{Operator/State Correspondence}
 
In the context of the large $N$ limit of Chern-Simons theory, no local
gravitation operators exist.  However in other applications, such as in the 
AdS/CFT context we are studying here, there are local deformations on both sides
and one needs to map them.  It is well known that the dictionary of AdS/CFT relates
a given state on the gravity side to an operator on the gauge theory side.  Here we would like
to comment that the fact that this is a one to one map fits naturally in the context of the \ova\
derivation of large $N$ duality.  Namely consider deforming the gravitational side by a vertex operator
corresponding to a scattering state.   In the limit of $t\rightarrow 0$ we ask where would the
vertex operator lie. It should be that they all lie in the Coulomb branch.  In other words
it should be that in the Higgs branch at $t =0$, cohomology is trivial.  Moreover it should
be that for each Coulomb branch region {\it at most one} vertex operator should be present deforming
the boundary conditions on the D-branes, which can be viewed as operator insertions from the
gauge theory perspective.
If there were non-vanishing contributions coming from 2 or more vertex operators on the Coulomb branch then the 1-1 correspondence between states and boundary operators will not work.  The approach of \ova\
is indeed compatible with this idea, because the A-model contribution
on the disc with no insertion
or 1 insertion is ambiguous (because there are residual conformal symmetries).  So in the computation of
$\oint dF $ they could lead to contributions.  However, for 2 or more insertions of vertex operators
in the C-region, $F$ is well defined and thus $\oint dF =0$.

With this review we are now ready to study the case of interest in this paper and see how much of this
structure carries over.

\newsec{Gauged Linear Sigma Model and Zero Radius Limit}

In proving the open-closed duality which relates $d=3$ Chern-Simons
theory and the resolved conifold, it was useful to write the topological
A-model for the resolved conifold as a gauged linear sigma model where
the parameter $t$ multiplying the A-model action is the Fayet-Illiopoulos
modulus in the gauged linear sigma model.
In the limit where $t\to 0$, the gauged linear sigma model for the
resolved conifold can develop both a Coulomb phase and a Higgs phase, 
and the Coulomb phase was interpreted as D-brane holes which
describe the faces in Feynman
diagrams of the Chern-Simons gauge theory.

In this section, we suggest that a similar technique may be useful for
obtaining
a worldsheet derivation of the Maldacena conjecture which relates
${\cal N}=4$ d=4 super-Yang-Mils and the $AdS_5\times S^5$ sigma model. 
In section (5.1), we shall write the A-model action of \sone\ as a
gauged linear sigma model with a $U(4)$ worldsheet gauge field. 
And in section (5.2), we shall argue that in the limit where $t\to 0$,
a Coulomb phase develops which can be interpreted as D-brane holes.
Furthermore, it will be argued that these D-brane holes are associated
with gauge-invariant ${\cal N}=4$ d=4 super-Yang-Mills operators.
Finally, a possible connection with the twistorial formulation of 
${\cal N}=4$ d=4 super-Yang-Mills will be discussed in subsection (5.3).

\subsec{Gauged linear sigma model}

The A-model action of \sone\ is based on the coset ${U(2,2|4)}\over
{U(2,2)\times U(4)}$ which can be interpreted as a ``fermionic'' version
of the Grassmannian ${U(M+N)}\over{U(M)\times U(N)}$. As shown in 
\ref\rocek{U. Lindstrom and M. Ro\v cek, {\it Scalar-Tensor Duality and
N=1, N=2 Non-Linear Sigma Models}, Nucl. Phys. B222 (1983) 285.}
\ref\cecotti{S. Cecotti and C. Vafa, {\it On Classification of N=2
Supersymmetric Theories}, Comm. Math. Phys. 158 (1993) 569, hep-th/9211097.},
the nonlinear sigma model action based on this Grassmannian can be written
as a gauged linear sigma model by introducing either a $U(M)$ or
a $U(N)$ worldsheet gauge field, together with an appropriate set
of matter fields transforming in the fundamental representation
of the gauge group. For the coset ${U(2,2|4)}\over {U(2,2)\times U(4)}$,
we shall choose to introduce a $U(4)$ worldsheet gauge field, however,
we suspect that 
the alternative choice of introducing a $U(2,2)$ gauge field would
not affect our conclusions.

In two-dimensional N=(2,2) superspace, the $U(4)$ worldsheet gauge field
is described by the real prepotential, 
$$V^R_S (z,\bar z,\k^+,\k^-,\bar\k^+,\bar\k^-),$$
where $R,S=1$ to 4 are local $U(4)$ indices, and the matter fields are
described by the chiral and antichiral superfields,
$$\Phi^\S_R(z,\bar z, \k^+,\k^-), \quad
\bar\Phi_\S^R(z,\bar z,\bar\k^+,\bar\k^-),$$
where $\S=(A,J)$ is a global $U(2,2|4)$ index and, as in the previous sections,
$A=1$ to 4 is a global $U(2,2)$ index and $J=1$ to 4 is a global $U(4)$ index.
Note that the matter fields transform in the fundamental representation
of the gauge group
and that $\Phi^A_R$ is a fermionic superfield whereas $\Phi^J_R$
is a bosonic superfield.

The gauged linear sigma model action is easily written in 
$U(2,2|4)$-invariant notation as 
\eqn\glsm{S = \int d^2 z \int d^4\k [ \bar\Phi_\S^S (e^V)^R_S \Phi^\S_R -
t V^R_R ]}
where $t$ is a constant parameter multiplying the Fayet-Illiopoulos term.
When $t$ is nonzero, this action is easily shown to be equivalent to
the A-model action of \sone\ by solving the equations of motion for
the preprotential $V_R^S$.
The equation of motion for $V^R_S$ is
\eqn\superv{t\d^R_S = (e^V)^R_T \bar\Phi^T_\S \Phi^\S_S,}
which implies that 
$$V^R_S = \d^R_S \log t - \log (\bar\Phi^R_\S \Phi^\S_S).$$

Plugging this auxiliary equation of motion into \glsm, one finds
\eqn\glsmn{S = t \int d^2 z \int d^4\k  Tr [\log (\bar\Phi_\S^R \Phi_S^\S)].}
Assuming that $\Phi_R^J$ and $\bar\Phi^R_J$ are invertible matrices,
one can define the chiral and antichiral superfields
$\T^A_J$ and $\Tb^J_A$ as 
\eqn\dettb{\T^A_J\equiv \Phi_R^A (\Phi_R^J)^{-1},\quad
\Tb^J_A\equiv \bar\Phi^R_A (\bar\Phi^R_J)^{-1},}
and write the action as
\eqn\glsmp{S = t \int d^2 z \int d^4\k  Tr [\log (\d^J_K +\bar\T^J_A \T^A_K)
+ \log(\Phi^J_R) + \log(\bar\Phi^R_J)].}
Since $\log(\Phi^J_R)$ is chiral and $\log(\bar\Phi^R_J)$ is antichiral,
the second and third terms of \glsmp\ vanish and the action coincides
with \sone.

\subsec{Zero radius limit}

As shown in \ref\phases{E. Witten,
{\it Phases of N=2 Theories in Two Dimensions}, Nucl. Phys. B403 (1993)
159, hep-th/9301042.}, the gauged linear sigma model is very convenient for
studying the limit where $t\to 0$. Since $t$ is identified with $\half R^2$
where $R$ is the $AdS_5\times S^5$ radius, this limit corresponds
to small 't Hooft coupling where perturbative ${\cal N}=4$ d=4 super-Yang-Mills
is a good description of the theory. In this $t\to 0$ limit, it
will be shown that the closed string variables in the gauged linear
sigma model can exist either in the Higgs phase where the $U(4)$
gauge symmetry is broken, or in the Coulomb phase where the $U(4)$
gauge symmetry is unbroken.

To analyze the different phases, 
we shall focus on the worldsheet fields with zero
conformal weight since only these fields can obtain nonzero
expectation values. After performing an A-twist, the only field
with zero conformal weight in
the prepotential $V_R^S$ is the complex field $\sigma_R^S$ 
where, in Wess-Zumino gauge, 
\eqn\vez{V_R^S = \s_R^S \k_+\bar\k_- + \bar\s_R^S \k_-\bar\k_+ + 
(A_z)_R^S \k_+\bar\k_+
+ (A_{\bar z})_R^S \k_-\bar\k_- + ... .}
And after an A-twist, the only matter fields with zero conformal weight
are $(\phi_R^\S,\psi_R^\S)$ and $(\bar\phi^R_\S, \bar\psi^R_\S)$ where
\eqn\phiez{\Phi_R^\S = \phi_R^\S + \k_+\psi_R^\S + ...,\quad
\bar\Phi^R_\S = \bar\phi^R_\S + \bar\k_-\bar\psi^R_\S + ... .}
Note that $(\phi_R^J,\psi_R^A,\bar\phi^R_J,\bar\psi^R_A)$ are bosonic
fields and
$(\phi_R^A,\psi_R^J,\bar\phi^R_A,\bar\psi^R_J)$ are fermionic fields.

If one sets to zero all component fields with nonzero conformal weight,
the equation of motion of \superv\ for $V_R^S$
implies that the remaining fields satisfy
\eqn\comp{\phi_R^\S \bar\phi_\S^S = \d_R^S t, \quad
\phi_R^\S \bar\psi_\S^S =0,\quad \psi^\S_R \bar\phi_\S^S=0,\quad
\psi_R^\S\bar\psi^S_\S = t \s_R^S.}

When $t$ is nonzero, the first equation of 
\comp\ implies that one can gauge $\phi_R^J = \d_R^J \sqrt{t}$
up to terms which are quadratic in the fermionic fields $\phi_R^A$.
The second and third equations of \comp\ define the fermionic fields
$\psi_R^J$ and $\bar\psi^R_J$ in terms of 
$(\phi_R^A,\psi_R^A)$ and $(\bar\phi^R_A,\bar\psi^R_A)$. And
the fourth equation implies that $\s_R^S$ is fixed to satisfy
$\s_R^S = t^{-1} \psi_R^\S \bar\psi^S_\S$. Since $U(4)$ symmetry
is broken and $\s_R^S$ is fixed, this is the Higgs phase.
In this phase, the unconstrained variables $(\phi_R^A,\psi_R^A)$ and
$(\bar\phi^R_A,\bar\psi^R_A)$ are related to the nonlinear sigma
model variables $(\t^A_J,Z^A_J)$ and $(\tb^J_A,\bar Z^J_A)$ using
the identification of \dettb.

When $t\to 0$, one possibility is that the worldsheet variables stay
in the Higgs phase. 
In this phase, $(\phi_R^A,\psi_R^A)$ and
$(\bar\phi^R_A,\bar\psi^R_A)$ are unconstrained, and
$\s_R^S$ is constrained to satisfy
$\s_R^S = t^{-1} \psi_R^\S \bar\psi^S_\S$ (which generically will diverge).
However, another possibility when $t\to 0$ is that $\s_R^S$
is unconstrained, but the matter variables are constrained to satisfy
\eqn\compt{\phi_R^\S \bar\phi_\S^S = 0, \quad
\phi_R^\S \bar\psi_\S^S =0,\quad \psi^\S_R \bar\phi_\S^S=0,\quad
\psi_R^\S\bar\psi^S_\S = 0.}
This phase will be called the Coulomb phase since the Cartan subgroup of $U(4)$ symmetry
is unbroken and $\s_R^S$ is unconstrained.  The fields $\phi_R^\S ,\psi^\S_R$ are massive
in this branch and can be integrated out.
If one Wick-rotates the $U(2,2|4)$ signature to a $U(4|4)$ signature,
then the equations of \compt\ imply that 
\eqn\dbr{\phi_R^\S = \psi_R^\S = \bar\phi^R_\S = \bar\psi^R_\S=0.}
This solution is analogous to the Coulomb phase for the resolved conifold
where all the matter fields are forced to vanish. So if one works
in the Euclidean signature of $U(4|4)$, the regions in the Coulomb
phase appear as ``holes'' in the closed string worldsheet. As in \ova ,
it is natural to identify the boundary of these holes with open
string D-branes
corresponding to the ``faces'' of the gauge theory
Feynman diagrams.

However, unlike in the conifold/Chern-Simons duality, one expects in the
AdS/Yang-Mills duality that
regions in the Coulomb phase 
carry additional information corresponding to the different
gauge-invariant super-Yang-Mills operators.
In other words, the D-brane holes can be described as boundary
states in the closed string theory, and the physical
closed string vertex operators for these boundary
states should correspond to gauge-invariant super-Yang-Mills operators.
As will now be shown, if one constructs
solutions to 
the Coulomb phase equations of \compt\ using
the original Minkowski space signature of $U(2,2|4)$, one can easily
describe 
the half BPS super-Yang-Mills operators.   This corresponds to the special
case where the boundary conditions on the D-brane have no spatial derivative.
It is still an open
question how to describe the non-BPS super-Yang-Mills operators.

To construct solutions to \compt\ in Minkowski space signature, it
will be convenient to split the $SU(2,2)$ index $A$ 
as $A=(a,\ad)$ where $a,\ad=1$ to 2, and to split the $SU(4)$ index $J$ as
$J=(j,j')$ where $j,j'=1$ to 2.
Furthermore, define $\bar\Phi^J_A$ to be the ``harmonic'' conjugate
of $\Phi^A_J$ where ``harmonic conjugation'' switches the $a$ and $\ad$
representations and also switches the $j$ and $j'$ representations. Note
that harmonic conjugation is equivalent to complex conjugation multiplied
by a $Z_2$ transformation in $SU(4)$, and is commonly used for
defining superfields in harmonic superspace.

With this definition of harmonic conjugation,
it is easy to see that 
\eqn\firs{\phi_R^j = \psi_R^j = \phi_R^a = \psi_R^a =0,\quad
\bar\phi^R_{j'} = \bar\psi^R_{j'} = \bar\phi^R_\ad = \bar\psi^R_\ad =0,}
is a solution of \compt\ which breaks $U(2,2|4)$ invariance to a
$U(1,1|2)\times U(1,1|2)$ subgroup. By deforming the solution of \firs\
using the ${U(2,2|4)}\over
{U(1,1|2)\times U(1,1|2)}$ parameters 
\eqn\supar{[x^a_{\ad}, \t^a_{j'}, \tb_\ad^j, u^j_{j'}],}
one discovers that the most general solution of \compt\ is
\eqn\solg{\phi_R^j = u^j_{j'}\phi_R^{j'} + \tb^j_\ad\phi_R^\ad,
\quad \psi_R^j = u^j_{j'}\psi_R^{j'} + \tb^j_\ad\psi_R^\ad,}
$$\phi_R^a = \t^a_{j'}\phi_R^{j'} + x^a_\ad\phi_R^\ad,\quad
\psi_R^a = \t^a_{j'}\psi_R^{j'} + x^a_\ad\psi_R^\ad,$$
$$\bar\phi^R_{j'} = u^j_{j'}\bar\phi^R_j + \t^a_{j'}\bar\phi^R_a, \quad
\bar\psi^R_{j'} = u^j_{j'}\bar\psi^R_j + \t^a_{j'}\bar\psi^R_a, $$
$$\bar\phi^R_{\ad} = \tb^j_\ad\bar\phi^R_j + x^a_\ad\bar\phi^R_a, \quad
\bar\psi^R_{\ad} = \tb^j_\ad\bar\psi^R_j + x^a_\ad\bar\psi^R_a,$$
where $(x^a_\ad,u^j_{j'})$ and $(\t^a_{j'},\tb_\ad^j)$ are eight
bosonic and eight fermionic parameters, and
$(\phi_R^{j'}, \psi_R^{j'}, \phi_R^\ad,\psi_R^\ad)$ and
$(\bar\phi^R_{j'}, \psi^R_{j'}, \phi^R_\ad,\psi^R_\ad)$ are unconstrained.

The parameters of \supar\ are precisely the projective harmonic
superspace variables used in \ref\siegel{M. Hatsuda and W. Siegel,
{\it A New Holographic Limit of $AdS_5\times S^5$}, Phys. Rev. D67 (2003)
066005, hep-th/0211184.}
and \ref\howe{P. Heslop and P.S. Howe, {\it Chiral Superfields in
IIB Supergravity}, Phys. Lett. B502 (2001) 259, hep-th/0008047.} to describe 
${\cal N}=4$ d=4 super-Yang-Mills
operators. As shown in \siegel,
the $U(2,2|4)$ generators can be expressed in terms of these
parameters as 
\eqn\genp{M_U^{U'} = {\p\over{\p y^U_{U'}}},\quad 
M_U^V = y_{U'}^V {\p\over{\p y_{U'}^U}} - C \d_U^V, \quad
M_{U'}^{V'} = - y_{U'}^{V} {\p\over{\p y_{V'}^V}} + C \d_{U'}^{V'},}
$$M_{U'}^U = - y_{U'}^V y_{V'}^U {\p\over{\p y_{V'}^V}} + 2 y_{U'}^U C,$$
where $U= (a,j)$, $U' =(\ad, j')$, $y_{U'}^U = (x_\ad^a, \t_{j'}^a,\tb_\ad^j,
u_{j'}^j)$, and $C$ is a central charge which commutes with the
$U(2,2|4)$ generators. 
%In terms of the $SU(2,2|4)$ generators
%$(M^J_A, M^A_J, M^A_B, M^J_K)$, the central charge $C$ is defined by
%the commutator
%\eqn\centralc{[M^J_A, M^B_K] = \d_A^B M^J_K - \d^J_K M_A^B + \d^J_K \d_A^B C.}
Note that the $U(1,1|2)\times U(1,1|2)$ generators $M_U^V$ and
$M_{U'}^{V'}$ act linearly on the harmonic variables of \supar.

In the Coulomb phase, one cannot define \dettb\ and
there is no connection between
the linear sigma model variables and the pure spinor variables of
\sigmainv. So the only sensible definition of the BRST operator in
the Coulomb phase are the usual N=2 worldsheet supersymmetry generators 
of the gauged linear sigma model
which transform the fields of zero conformal weight as
\eqn\normalbrst{Q \phi_R^\S = \psi_R^\S,\quad \bar Q \bar\phi^R_\S = \bar\psi^R_\S.}
Although this BRST transformation preserves the full $U(2,2|4)$ invariance,
there is no contradiction when $t=\half R^2\to 0$ since, in this limit,
the super-Yang-Mills theory has no interaction terms which means it
contains the bonus $U(1)$ symmetry.

When the worldsheet variables are in the Coulomb phase, it is easy
to verify from the solution of \solg\ that 
$Q \phi_R^{j'} = \psi_R^{j'}$, 
$Q \phi_R^{\ad} = \psi_R^{\ad}$, 
$\bar Q \bar\phi^R_j = \bar \psi^R_j$, 
$\bar Q \bar \phi^R_a =\bar  \psi^R_a$, and
the harmonic variables of \supar\ are BRST invariant.
Therefore, any function of 
$[x^a_{\ad}, \t^a_{j'}, \tb_\ad^j, u^j_{j'}]$ which is independent
of $(\phi_R^{j'}, \psi_R^{j'}, \phi_R^\ad,\psi_R^\ad,
\bar\phi^R_{j'}, \psi^R_{j'}, \phi^R_\ad,\psi^R_\ad)$ 
is in the BRST cohomology.

But as was shown in \howe,
supergravity solutions on $AdS_5\times S^5$
with $C-2$ units of $S^5$ angular momentum
are in one-to-one corespondence with functions of 
$[x^a_{\ad}, \t^a_{j'}, \tb_\ad^j, u^j_{j'}]$ with a given central charge
$C$. So these functions in the BRST cohomology describe the
half BPS gauge-invariant super-Yang-Mills operators. For example, if
$V_C$ is a function with central charge $C$, then $V_C=1$ corresponds
to the super-Yang-Mills state $Tr (Z^C)$ where $Z$ is the scalar with
$U(1)$ charge $+1$ with respect to $M^j_j$ of \genp. The other half BPS
states with $C$ super-Yang-Mills fields can be obtained from $V_C=1$
by hitting with the generators $M_{U'}^U$ of \genp.

So at $t=0$ in the Coulomb phase, functions in the BRST cohomology
describe the half BPS gauge-invariant super-Yang-Mills operators at
zero coupling. Up to now, only constant modes of the worldsheet variables
in the Coulomb phase have been considered. It is plausible
that non-BPS gauge-invariant super-Yang-Mills operators will
be described by functions which also depend on non-constant
modes. This seems natural since half BPS operators are related to
``massless''
closed string vertex operators which only depend on constant modes, whereas
non-BPS operators are related to ``massive'' closed string vertex operators
which depend on non-constant modes of the worldsheet variables.

\subsec{Twistorial formulation}

It is quite intriguing that the gauged linear sigma model we are studying is closely
related to the twistor space relevant for the formulation of ${\cal N}=4$ Yang-Mills \ref\wittwis{
E. Witten, {\it Perturbative Gauge Theory as a String Theory in Twistor Space}, 
Comm. Math.  Phys. 252 (2004) 189, hep-th/0312171.
}.  In particular consider the limit $t\rightarrow 0$ and focus on the Coulomb branch of our $U(4)$ gauged linear sigma model in this limit.  A generic point on the Coulomb branch corresponds to breaking
the gauge group to $U(1)^4/S_4$ where $S_4$ is the permutation group on the
four $U(1)$ factors.  In this case the charged matter multiplet for each $U(1)$ corresponds
to a $(4|4)$ space.  Thus the gauged linear sigma model will give rise to the Coulomb branch of the topological
A-model of four copies of ${\bf CP}^{3|4}$, one for each $U(1)$.  In other words the corresponding non-linear
sigma model in its Higgs branch is
$$Sym^{\otimes 4}({\bf CP}^{3|4}).$$
Note that the above construction is related to the well-known geometric
fact that the
Grassmannian ${U(n+m)}\over{U(n)\times U(m)}$ can be viewed as
$Sym^{\otimes m}({\bf CP}^{m+n-1})$ \cecotti.

The appearance of the twistor space in our context is rather interesting and suggests
perhaps another view of our link to gauge theory.  In fact the twistor space does seem to play a role in the boundary conditions we have found
since, as discussed in \siegel, twistors and harmonic variables are related
when the fields are onshell. 
As explained in \siegel, the onshell equations for
a function $f(y)$ where 
$y^{U}_{U'}=[x^a_{\ad}, \t^a_{j'}, \tb_\ad^j, u^j_{j'}]$ are
\eqn\onshell{{\p\over{\p y^{[U}_{[U'}}}
{\p\over{\p y^{V]}_{V']}}} f(y)=0,}
which implies
the mass-shell condition ${\p\over{\p x^{a\ad}}} {\p\over{\p x_{a\ad}}} f(y)=0$.
The onshell equations of \onshell\ can be easily solved by writing the
Penrose-like transform
\eqn\twisto{f(y) = \int d\zeta \tilde f(\zeta_U, \zeta_{U'}\equiv \zeta_U y^U_{U'})}
where $(\zeta_U,\zeta_{U'})$ are ${\bf CP}^{3|4}$ twistor variables and
$\tilde f(\zeta)$ is a twistor function of the appropriate $U(1)$ weight.

However there is a difference between the appearance of twistors here and the one in \wittwis :  In
that case one was dealing with the topological B-model, whereas here we are dealing
with the topological A-model.  In fact the situation here is more similar to the setup 
considered in \ref\NV{A. Neitzke and C. Vafa, {\it N=2 Strings and the
Twistorial Calabi-Yau}, hep-th/0402128.}\
where the open A-model topological string on ${\bf CP}^{3|4}$ was proposed to be a perturbative
realization of ${\cal N}=4$ YM.  This could be related by S-duality to the formulation of \wittwis. 
It could also be that the 
A-model and B-model theory can appear similar in a hyperkahler setup
%NB removed the word "observable"
as is the case in \ref\kapw{A. Kapustin and E. Witten,
{\it Electric-Magnetic Duality and the Geometric Langlands Program},
hep-th/0604151.}.  It would be interesting to explore the connection with
twistor space further.

\newsec{Conclusions and Open Problems}

We have taken a step towards a worldsheet derivation of the Maldacena conjecture.  In particular
we have argued that the A-model topological string on ${U(2,2|4)}\over
{U(2,2)\times U(4)}$, which describes
string theory in the background of $AdS_5\times S^5$, is a gauged linear sigma model that in the
small radius limit develops a new branch, the Coulomb branch, which creates `holes' on the worldsheet.
This should correspond to the open string diagrams describing ${\cal N}=4$ supersymmetric Yang-Mills.
As evidence for this we showed that the half BPS operators of the gauge theory naturally arise
as solutions to the boundary conditions of the Coulomb branch.
  
  There
are a number of things that need to be better understood.  One has to analyze
the effect of integrating out the degrees of freedom on the Coulomb branch and show
that they give rise to the factor $N\lambda$ as is expected from the Chan-Paton factors.
More generally one would like to show, in addition to the half BPS states that we discussed,
how the precise dictionary between gravitational states and gauge theory operators work.

We have found an intriguing connection to (four copies of) the twistor space.  This is very
suggestive and calls for a deeper understanding of the role of twistors in the worldsheet derivation
of the Maldacena conjecture.

\vskip 1cm
{\bf Acknowledgment}

We would like to thank O. Aharony, R. Gopakumar, P. Howe, J. Maldacena, L. Motl, H.
Ooguri,
M. Ro\v cek, W. Siegel, B.C. Vallilo and E. Witten for valuable discussions.
We would also like to thank the 5th Simons Workshop in Mathematics and Physics 
and the Strings 2007 Conference for their hospitality.
The research of N.B. was supported in part by
CNPq grant 300256/94-9 and FAPESP grant 04/11426-0.
The research of C.V. was supported in part by NSF grants PHY-0244821 and DMS-0244464.

\listrefs

\end